\documentclass[12pt,epsf]{article}

\newlength{\pubnumber} 

\catcode`\@=11
\@addtoreset{equation}{section}

\def\section{\@startsection{section}{1}{\z@}{3.5ex plus 1ex minus .2ex}
 {2.3ex plus .2ex}{\large\bf}}
\def\subsection{\@startsection{subsection}{2}{\z@}{2.3ex plus .2ex}
 {2.3ex plus .2ex}{\bf}}

\usepackage{graphicx}
\usepackage{dcolumn}
\usepackage{bm}

\def\beq{\begin{equation}}
\def\eeq{\end{equation}}
\def\beqn{\begin{eqnarray}}
\def\eeqn{\end{eqnarray}}

\def\half{{\textstyle{1\over 2}}}

\begin{document}
\begin{titlepage}
\setcounter{page}{1}
\rightline{BU-HEPP-08-04}
\rightline{CASPER-08-01}
\rightline{\tt }

\vspace{.06in}
\begin{center}
{\Large \bf Radius Destabilization in Five Dimensional \\ 
            Orbifolds Due to an Enhanced Casimir Effect}\vspace{.12in}

{\large
        R. Obousy\footnote{Richard\_K\_Obousy@baylor.edu}
        and G. Cleaver\footnote{Gerald\_Cleaver@baylor.edu}}
\\
\vspace{.12in}
{\it        Center for Astrophysics, Space Physics \& Engineering Research\\
            Department of Physics, Baylor University,
            Waco, TX 76798-7316\\}
\vspace{.06in}
\end{center}

\begin{abstract}
One of the challenges in connecting higher dimensional theories to cosmology is stabilization of the moduli fields. 
We investigate the role of a Lorentz violating vector field in the context of stabilization. Specifically, we compute the one loop Casimir energy in Randall-Sundrum 5-dimensional (non-supersymmetric) $S^1/ Z_2$ orbifolds resulting from the interaction of a real scalar field with periodic boundary conditions with a Lorentz violating vector field. We find that the result is an enhanced attractive Casimir force. Hence, for stability, positive contributions to the Casimir force from branes and additional fields would be required to counter the destabilizing, attractive effect of Lorentz violating fields.
\end{abstract}
\end{titlepage}
\setcounter{footnote}{0}


\section{Introduction}

The proposal that there exists extra spatial dimensions in which gravity and possibly other fields can propagate has been the subject of intense study of recent \cite{aadd},\cite{add},\cite{s}. Of particular interest is the Randall-Sundrum (RS) model which addresses the heirachy problem \cite{rs},\cite{rs1}. In the RS setup the Standard model fields are confined to one of two 3-branes which lie at the endpoints (i.e., fixed points) of an $S^1/Z_2$ orbifold. The line element in RS is described by the metric
\begin{equation}
ds^2=e^{-2kR|\varphi|}\eta_{\mu\nu}dx^\mu dx^\nu - R^2 d\varphi^2,
\label{eq1}
\end{equation}
where the points $(x^\mu,\varphi)$ and $(x^\mu,-\varphi)$ are identified with each other, $x^\mu$ are the standard four dimensional coordinates and $|\varphi|\leq\pi$. The exponential factor is referred to as the warp factor and is an appealing feature in the RS model, as it can generate a TeV mass scale from the Planck scale in the higher dimensional theory, while retaining a bulk width that is only a couple of orders of magnitude above the Planck scale. A field with mass $m_0$ on the $\varphi=0$ brane will have a reduced physical mass of $m\approx \rm{e}^{-2\pi k R}m_0$ on the $\varphi=\pi$ brane. Typically $2\pi kR\approx 12$. In this model the branes themselves remain static and flat, and the fields confined to them preserves 4D Lorentz invariance. However no such restrictions need apply to fields extended along the extra dimension.

Four-dimensional Lorentz invariance is a basic ingredient in the standard model (and all local relativistic quantum field theories) which has been verified by numerous experiments \cite{k}. However, motivation does exist for deeper study into possible Lorentz violation. One reason is that quantitive statements regarding the degree with which nature preserves Lorentz symmetry are expressed within a framework which \textit{allows} for violations \cite{kp}. Another compelling reason is that the sensitivity of current tests implies that highly supressed Lorentz violations might arise at scales well beyond standard model physics.

It has been shown that spontaneous Lorentz breaking may occur in the context of some string theories \cite{ks}. In the standard model spontaneous symmetry breakdown occurs when symmetries of the Lagrangian are not obeyed by the ground state of the theory. This occurs when the perturbative vacuum is unstable. The same ideas apply in covariant string theory which, unlike the standard model, typically involve interactions that could destabilize the vacuum and generate nonzero expectation values for Lorentz tensors (including vectors) \cite{ck}.

A simple mechanism to implement local Lorentz violation is to postulate the existence of a tensor field with non-zero expectation value which couples to standard model fields. The most elementary realization of this is to consider a single spacelike vector field with a fixed norm. This field selects a `preferred' frame at each point in spacetime and any fields that couple to it will experience a local violation of Lorentz invariance.

Recently, Carroll \cite{ct} investigated the role of Lorentz violating fields in hiding extra dimensions. One novel feature of the research was the demonstration that it allowed different spacings in the Kaluza Klein towers. The model worked in a five dimensional flat spacetime, and the Lorentz violating field is a spacelike five-vector $u^a=(0,0,0,0,v)$, which ensures four dimensional Lorentz invariance is preserved. The fifth dimension is compactified on a circle. If we first define an antisymmetric `Lorentz Violation Tensor' $\xi^{ab}$  in terms of $u^a$

\begin{equation}
\xi_{ab}=(\nabla_a u_b -\nabla_b u_a)
\label{eq3},
\end{equation}
we can form the following action: 
\begin{equation}
S=M_*\int d^5x \sqrt{g} \left[-\frac{1}{4}\xi_{ab}\xi^{ab}-\lambda(u_au^a-v^2)+\sum_{i=1} \mathcal{L}_i\right].
\label{eq2}
\end{equation}
Here the indices $a, b$ run from 0 to 4. $\lambda$ is a Lagrange multiplier which ensures $u^au_a=v^2$, and we take $v^2>0$. The $\mathcal{L}_i$ can represent various interaction terms. For this letter we will only investigate interactions with a scalar field. This form of the Lagrangian ensures the theory remains stable and propagates one massless scalar and one massless pseudoscalar \cite{dgb}. Of interest is the Kaluza-Klein tower generated by the Lorentz violating field in the compact dimension in the context of moduli stabilization. 

There has been a huge body of work in the context of the supergravity limits of superstring theory to stabilize the moduli fields. One popular approach is to introduce fluxes about the compactified spaces to stabilize the shape moduli \cite{s86,wsd}. An alternative approach arises in string gas cosmology which is based on coupling a gas of strings to some standard background \cite{cr}. In this process it is the winding and momentum modes of strings which contributes a negative pressure term and thus play an important role in the dynamics of the extra dimensions.

Another popular mechanism employed in the study of stabilization is the Casimir effect  \cite{i}, \cite{pp}, \cite{enoo}, \cite{cenoz}. In its simplest form, the Casimir effect is the interaction of a pair of neutral, parallel conducting planes whose existence modify the ground state of the quantum vacuum in the interior portion of the plates creating a force which attracts the plates to each other. For a review on the Casimir effect see \cite{m}, \cite{bmm}.  

The Casimir effect can be extended to regions of non-trivial topology \cite{aw}, \cite{ke}. For example, on $S^1$, a circular manifold, one can associate $0$ and $2\pi$  with the location of the plates and the Casimir energy can be calculated. This becomes relevant when we consider models with additional spatial dimensions  \cite{gl},  \cite{bcp}. The Casimir force has also been studied for the case of a Lorentz violating theory where it was discovered that the force was modified \cite{ft}, and in the context of the Randall Sundrum model where the corrections due to the extra dimension was quantified \cite{ftz}.

Clearly all fields which propagate in the bulk will give Casimir contributions to the vacuum energy and a natural extension of the study of Lorentz violating fields is whether these could provide a stabilizing force. In this letter, we calculate the effective potential due to a Lorentz violating tensor field coupling with a scalar field with periodic boundary conditions. We will focus on the background geometry of the Randall-Sundrum model, although the techniques employed here may be used in alternative geometries.

This letter will be organized as follows. In section 2 we will begin by reviewing the Casimir effect. In section 3 we review the calculations involved in acquiring the KK tower of a scalar field coupled to a Lorentz violating spacelike five vector, and finally in section 4 we derive the Casimir energy of this scenario using zeta function regularization addressed within the context of moduli stability, which is the main result of this letter.   	 	

\section{The Casimir Effect}

Arguably the most poignant demonstration of the reality of the quantum vacuum is the famous Casimir effect. In 1948 H. Casimir published a profound paper where he explained the van der Waals interaction in terms of the zero-point energy of a quantized field \cite{cas}. In its most elementary form the Casimir effect is the interaction of a pair of neutral parallel plates. The presence of the plates modifies the quantum vacuum and this modifcation causes the plates to be pulled toward each other with a force $F\propto{a^{-4}}$, where a is the plate separation. For many years the paper remained unknown \cite{m} but from the 70's onwards the Casimir effect received increasing attention and in the last decade has become very popular. 

One intriguing aspect of the Casimir effect is that it is a purely quantum effect. In classical electrodynamics the force between the two parallel plates is zero. However, in the ideal scenario at zero temperature (where there are no real photons between the plates, only virtual photons) it is the ground state of the quantum electrodynamic vacuum which causes the attraction. The most important feature of the Casimir effect is that even though it is purely quantum in nature, it manifests itself macroscopically. For example, for two plates of area $A=1cm^2$ separated by a distance of $d=1\mu m$ the force of attraction is $F \approx 1.3 \times 10^{-7}N$. This force is certainly within the range of laboratory force measuring techniques. Something that is unique to the Casimir force is that both the sign and the magnitude of the Casimir force is strongly dependant on the geometry of the plates. This makes the Casimir effect a good candidate for applications in nanotechnology \cite{bmm}.

Typically the calculations of vacuum expectation values (VEV's) are divergent so some from of renormalization must be performed. For example, consider the calculation of the VEV's inside a metal cavity. Such a calculation will necessarily involve summing the energies of the standing waves in the cavity to infinity.
\begin{equation}
\left<E_{vac}\right>=\frac{1}{2}\sum_{n=1}^\infty E_n
\label{eq3}
\end{equation}
It is for this reason that a variety of sophisticated mathematical procedures must be employed so as to extract finite quantities. Examples of these procedures include introducing physical cutoffs, dimensional regularization and Green's function techniques.

In the last few years the study of the Casimir effect has become increasingly popular and is studied in the context of a wide variety of field in physics including Gravitation and Cosmology, Condensed Matter Physics, Atomic and Molecular Physics, Quantum Field Theory and even Nanotechnology \cite{m}. Recently, high precision experiments have been performed demonstrating the Casimir force and more experiments are on the way. 

\subsection{Casimir's Original Calculations}

In this section we review Casimir's original calculations \cite{cas} and the approach he took to controlling the divergences associated with zero-point energy. The calculation will be performed for a massless scalar field in one dimension. The setup is as follows. We place a perfectly conducting plate at $x=0$ and $x=L<M$, where $M$ will eventually go to $\infty$. In this setup the energy spectrum is discrete in the region $0<x<L$. To model the plates we impose the following boundary conditions on the field:
\begin{equation}
\phi(t,0)=\phi(t,L)=0 \ \ \ \ \forall \ \ t.
\label{eq4}
\end{equation}
Using these conditions we can determine the allowed modes, k. Due to the boundary conditions the only possible modes are
\begin{equation}
k_n=\frac{n\pi}{L}.
\label{eq5}
\end{equation}
The total energy is given by summing over all modes.
\begin{equation}
E_0(L)=\frac{\pi}{2L}\sum_{n=0}^\infty n
\label{eq6}
\end{equation}
This sum is clearly divergent. We can also calculate the energy density on the other side of the plate, e.g for $x>L$
\begin{equation}
\mathcal{E}_0=\lim \limits_{M \to \infty}\frac{1}{M-L}E_0(M-L)= \lim \limits_{\Delta x \to \infty}\frac{\pi}{2} \sum_{n=0}^\infty(n\Delta x)\Delta x=\frac{\pi}{2}\int_0^\infty xdx.
\label{eq7}
\end{equation}
Between the plates it is clear that we are faced with an infinite sum. This is because the modes are discretized due to the boundary conditions imposed by the presence of the plates. Exterior to the plates no such boundary conditions exist and so the modes become continuous and the sum turns into an integral. Clearly the total energy stored in the field in the region $0\to M$ is the sum of the above contributions.
\begin{equation}
E_{tot}(L)=E_0(L)+\lim \limits_{M \to \infty}(M-L)\mathcal{E}_0
\label{eq8}
\end{equation}
Note that both $E_0(L)$ and $\mathcal{E}_0$ are infinite quantities. To define these in a mathematically valid sense we introduce an exponential cutoff parameter $\alpha>0$ regularizing the energies.
\begin{equation}
E_0^{reg}(L)=\frac{\pi}{2}\sum_{n=1}^\infty\frac{n}{L}e^{-\frac{\alpha n}{L}}
\label{eq9}
\end{equation}
The original series is recovered in the limit $\alpha \to 0$. We can rewrite the above equation in terms of a derivative.
\begin{equation}
E_0^{reg}(L)=\frac{\pi}{2} \frac{\partial}{\partial \alpha}\sum_{n=1}^\infty e^{-\frac{\alpha n}{L}}
\label{eq10}
\end{equation}
For $\alpha>0$ this term is finite. Observe that the denominator $\to 0$ as $\alpha \to 0$. A physical interpretation of this exponential cutoff is that any material becomes transparent at high enough frequencies. Eq.\ (\ref{eq10}) can be expanded as a Taylor series so that we can observe the source of the divergence:
\begin{equation}
E_0^{reg}(L)=\frac{\pi}{2\alpha^2}L-\frac{\pi}{24L}+ \frac{\pi}{480L^3}\alpha^2+\mathcal{O}(\alpha^4)
\label{eq11}
\end{equation}
Clearly for $\alpha=0$ a pole of order 2 exists as indicated by the first term in the equation. Having introduced the cutoff for the summation in eq.\ (\ref{eq7}). we can apply the same process to the integral in eq.\ (\ref{eq7}): 
\begin{equation}
\mathcal{E}_0=\frac{\pi}{2}\int_0^\infty x e^{-\alpha x}dx=\frac{\pi}{2\alpha^2}
\label{eq12}
\end{equation}
Because $E_0^{tot}(L)$ depends on the position of the plate, there must be a force on it. This can be found by taking the derivative of the energy with respect to the plate distance:
\begin{equation}
\mathcal{F}(L)=-\frac{\partial}{\partial L}E_0(L)+\mathcal{E}_0=-\frac{\partial}{\partial L}(\frac{\pi}{2\alpha^2}L-\frac{\pi}{24L}+ \mathcal{O}(\alpha^2))+\frac{\pi}{2\alpha^2}
\label{eq13}
\end{equation}
Giving us the result:
\begin{equation}
\mathcal{F}_{\alpha}=-\frac{\pi}{24L^2}+\mathcal{O}(\alpha^2)
\label{eq14}
\end{equation}
Now as we take $\alpha \to 0$ we obtain the Casimir force:
\begin{equation}
\mathcal{F}=-\frac{\pi}{24L^2}
\label{eq15}
\end{equation}
This remarkable result demonstrates that even in empty space in the absence of any external forces, there exists a force of attraction between two parallel plates, the origin of which is purely quantum theoretic.

\subsection{\bf Alternative Derivations}

In the following sections two sections we briefly review alternative derivations of the Casimir energy.

\subsubsection{\bf Riemann Zeta Function}

There is a remarkable simplicity to the Casimir calculation when the Riemman zeta function is utilized. Recall the expression of the energy between the plates:
\begin{equation}
E_0(L)=\frac{\pi}{2L}\sum_{n=1}^\infty n
\label{eq16}
\end{equation}
If we use the definition of the Riemann zeta function:
\begin{equation}
\zeta(s)=\sum_{n=1}^\infty \frac{1}{n^s}
\label{eq17}
\end{equation}
we can rewrite eq.\ (\ref{eq16}) as:
\begin{equation}
E_0(L)=\frac{\pi}{2L}\sum_{n=1}^\infty \frac{1}{n^{-1}}=\frac{\pi}{2L}\zeta(-1)
\label{eq18}
\end{equation}
However, $\zeta(-1)=-\frac{1}{12}$ from analytic continuation and so we quickly obtain the result:
\begin{equation}
E_0(L)=-\frac{\pi}{24L}
\label{eq19}
\end{equation}
and next take the derivative to obtain the force:
\begin{equation}
\mathcal{F}(L)=-\frac{\partial}{\partial L}E_0(L)=-\frac{\pi}{24L^2}
\label{eq20}
\end{equation}	
which is in agreement with eq.\ (\ref{eq15}).

\subsubsection{\bf Analytic Continuation}

Another technique used to control the divergences associated with Casimir energy calculations involves utilizing analytic continuation as described in \cite{aw} and \cite{thv} for example. Again, we consider a scalar field that satisfies the free Klein Gordon equation in d dimensions.
\begin{equation}
(\partial^2+m^2)\phi(x)=0
\label{eq21}
\end{equation}
in the absence of boundaries. Constraining the fields as $x=0$ and $x=a$ we impose Dirichlet boundary conditions, e.g.
\begin{equation}
\phi(0)=\phi(a)=0
\end{equation}
The modes of this field are then
\begin{equation}
\phi(x,t)=sin(\frac{n\pi x}{a})e^{i({\bf k.x}-\omega_k t)}
\label{eq22}
\end{equation}
or,
\begin{equation}
\omega_k=\sqrt{\frac{n\pi}{a}+{\bf k}^2+m^2}
\label{eq23}
\end{equation}
where n is a positive integer. In the ground state, each mode contributes an energy $\frac{1}{2}\omega_k$. The total energy of the field between the plates is
\begin{equation}
E=\frac{L}{2\pi}^{d-1}\int d^{d-1}{\bf k}\sum_{n=1}^\infty \frac{1}{2}\omega_k
\label{eq24}
\end{equation}
The sum is clearly divergent but can be regularized by using a process of analytic continuation. Using the forumla
\begin{equation}
\int d^dk f(k)=\frac{2\pi^{d/2}}{\Gamma(\frac{d}{2})}\int k^{d-1}f(k)dk
\label{eq25}
\end{equation}
and substituting into eq.\ (\ref{eq24}), we obtain
\begin{equation}
E=\left(\frac{L}{2\pi}\right)^{d-1}\frac{2\pi^{(d-1)/2}}{\Gamma\frac{d-1}{2}}\sum_{n=1}^\infty\int_0^\infty \frac{1}{2}({\bf k}^2)^{(d-3)/2}d({\bf k}^2)\frac{1}{2}\sqrt{\left(\frac{n\pi}{a}\right)^2+{\bf k}^2+m^2  }
\label{eq26}
\end{equation}
Using the well known expression for the Beta function,
\begin{equation}
\int_0^\infty t^r(1+t)^s dt=B(1+r,-s-r-1)
\label{eq27}
\end{equation}
and plugging in for the Beta function, we obtain
\begin{equation}
E=\frac{1}{2}\frac{\Gamma\left(\frac{-d}{2}\right)}{\Gamma\left(\frac{-1}{2}\right)}\pi^{(d+1)/2}\frac{(L/2)^{d-1}}{a^d}\sum_{n=1}^\infty\left[\left(\frac{am}{\pi}\right)^2+n^2\right]^{d/2}
\label{eq28}
\end{equation}
At this stage we can introduce the Riemann zeta function, as the sum is clearly divergent. Also, utilizing the reflection formula
\begin{equation}
\Gamma\frac{s}{2}\pi^{-s/2} \zeta(s)=\Gamma\left(\frac{1-s}{2}\right)\pi^{(s-1)/2}\zeta(1-s)
\label{eq29}
\end{equation}
and the reduplication formula
\begin{equation}
\Gamma(s)\sqrt{\pi}=2^{s-1}\Gamma\left(\frac{s}{2}\right)\Gamma\left(\frac{1+s}{2}\right)
\label{eq30}
\end{equation}
we can re-write the energy as
\begin{equation}
E=-\frac{L^{d-1}}{a^d}\Gamma\left(\frac{d+1}{2}\right)(4\pi)^{-(d+1)/2}\zeta(d+1)
\label{eq31}
\end{equation}
The result is finite for all d and always negative.  For the case of d=1 we obtain
\begin{equation}
E=-\frac{\pi}{24a}
\label{eq32}
\end{equation} 
Again, the force is obtained by taking the derivative,$-\frac{\partial(E/L^{d-1})}{\partial a}$ and the result is in agreement with eq.\ (\ref{eq15}).

\section{KK Spectrum for Periodic Scalars Interacting with Lorentz Violating Vectors}

Having reviewed the Casimir effect we will now determine the KK spectrum for periodic scalars interacting with Lorentz violating vectors. We consider a real scalar field $\phi$ coupled to a Lorentz violating spacelike five vector $u^a$ with a VEV in the compact extra dimension. The Lagrangian is
\begin{equation}
\mathcal{L}_{\phi}=  \frac{1}{2} (\partial\phi)^2 -\frac{1}{2} m^2\phi^2-\frac{1}{2\mu^2_\phi}u^a u^b\partial_a\phi\partial_b\phi.
\label{eq33}
\end{equation}
The indices a and b run from 0 to 4. The mass scale $\mu_\phi$ is added for dimensional consistency. The background solution has the form $u^a=(0,0,0,0,v)$ which ensures four dimensional Lorentz invariance is preserved. 

Using the five dimensional Euler Lagrange equation
\begin{equation}
\partial_a \left(\frac{\partial \mathcal{L}}{\partial(\partial_a\phi)}\right)-\frac{\partial \mathcal{L}}{\partial \phi}=0
\label{eq34a}
\end{equation}
and plugging in for the Lagrangian we obtain
\begin{equation}
\partial_a\partial^a\phi-m^2\phi=\mu^{-2}_\phi\partial_a(u^au^b\partial_b\phi).
\label{eq34b}
\end{equation}
The scalar can be expressed in momentum space as
\begin{equation}
\phi \propto e^{ik_ax^a}=e^{ik_\mu x^\mu}e^{ik_5y}
\label{eq35}
\end{equation}
where $\mu=0,1,2,3$. Calculating each term in the Euler Lagrange equation (\ref{eq34b}),
\begin{eqnarray}
\partial_a\partial^a\phi &=&\partial_\mu\partial^\mu\phi+\partial_y\partial^y\phi
\label{eq36}\\ 
                         &=&  -k_\mu k^\mu\phi-k_5 k^5\phi .
\label{eq37} 
\end{eqnarray}
For the term involving the VEV of the Lorentz violating field it is clear that the only nonzero index values are $a=b=5$, thus we quickly obtain
\begin{eqnarray}
\mu^{-2}_\phi\partial_a(u^au^b\partial_b\phi)
&=&\mu^{-2}_\phi\partial_5(u^5u^5\partial_5\phi) \label{eq38}\\ 
&=&  \frac{v^2k_5^2}{\mu_\phi^2}\phi \label{eq39}. 
\end{eqnarray}
where we have used the fixed norm constraint $u^au_a=v^2$ obtained from the equation of motion for $\lambda$. Choosing $v^2>0$ ensures that the vector will be timelike. We now compactify the fifth dimension on a circle of radius R ($k_5=\frac{n\pi}{R}$), with $Z_2$ symmetry which identifies $u^a\rightarrow -u^a$. 

The orbifolding will not effect the coupling of the \textit{scalar} field to the Lorentz violating field, but would effect the couplings for more complex lagrangians (fermions for example) and so we include this precedure for completeness, and also for its relevance in the RS paradigm. The effect of the orbifolding is essentially to remove all odd (even) scalar modes under $y\rightarrow -y$ for even (odd) periodicity scalar fields. In each case this amounts to eliminating half of the modes in summation over $n$ \cite{pp}. Thus, for both periodicities, the net result is a reduction of the Casimir energy be a factor of $\half$.

Plugging eq.\ (\ref{eq37}) and eq.\ (\ref{eq39}) into eq.\ (\ref{eq34a}) we obtain
\begin{equation}
-k_\mu k^\mu=m^2+(1+\alpha_\phi^2)\left(\frac{n\pi}{R}\right)^2
\label{eq40}
\end{equation}
where $\alpha_\phi=\frac{v}{\mu_\phi}$ is the ratio of the Lorentz violating VEV to the mass parameter.

We thus see that as shown in \cite{ct} with the addition of a Lorentz violating field the mass spectrum of the extra dimensional Kaluza Klein tower is modified by non-zero $\alpha_\phi$: 
\begin{equation}
m^2_{KK}=k^2+(1+\alpha_\phi^2) \left( \frac{n\pi}{R} \right) ^2
\label{eq41}
\end{equation}
The value of $\alpha_\phi$ depends on the choice of the mass scale which should be on the order of the Planck scale.

\section{Moduli Stability}

We now apply the results of \cite{ct} reviewed in the previous section to the question of moduli stabilization. The Casimir energy due to the Kaluza-Klein modes of a scalar field obeying periodic boundary conditions compactified on $S^1$ and interacting with a Lorentz violating vector field is  
\begin{equation}
E=\frac{1}{2}{\sum_{n=-\infty}^{\infty}}'\int\frac{d^4k}{(2\pi)^4}log\left( k^2+(1+\alpha_\phi^2)\left(\frac{n\pi}{R} \right)^2         \right), 
\label{eq42}
\end{equation}
where the prime on the summation indicates that the $n=0$ term is omitted. We can rewrite the log as a derivative and then after a Mellin transform perform a dimensional regularization on the integral and the summation
\begin{eqnarray}
E &=&\frac{1}{2}\frac{\partial}{\partial s}|_{s=0}{\sum_{n=-\infty}^{\infty}}'\int\frac{d^4k}{(2\pi)^4}\left( k^2+\xi n^2 \right)^{-s} 
\label{eq43}\\
  &=&\frac{1}{2}\frac{\partial}{\partial s}\zeta^+(s) |_{s=0}
  \label{eq44},
\end{eqnarray}
where the periodic scalar function is defined as
\begin{equation}
\zeta^+(s)={\sum_{n=-\infty}^{\infty}}'\int\frac{d^4k}{(2\pi)^4}\frac{1}{\Gamma(s)}\int_0^\infty dte^{(k^2+\xi n^2)t}t^{s-1}. 
\label{eq45}
\end{equation}
Here we have made the substitution $\xi=\frac{\pi^2(1+\alpha^2_\phi)}{R^2}$ and used the identity
\begin{equation}
z^{-s}=\frac{1}{\Gamma(s)}\int_0^\infty dt e^{-zt}t^{s-1}.
\label{eq46}
\end{equation}

We first perform the k integral 
\begin{equation}
\int d^4k e^{-k^2t}=\frac{\pi^2}{16t^2},
\label{eq47}
\end{equation}
and now calculate
\begin{equation}
\zeta^+(s)=\frac{\pi^2}{(2\pi)^4}\frac{1}{\Gamma(s)}{\sum_{n=-\infty}^{\infty}}'\int_0^\infty dt e^{-\xi n^2t}t^{s-3}.
\label{eq48}
\end{equation}
Making the substitution $x=\xi n^2 t$ gives us
\begin{eqnarray}
t=\frac{x}{\xi n^2}\quad{\rm and}\quad dt=\frac{dx}{\xi n^2}.
\label{eq49}\end{eqnarray}
Now substituting back into eq.\ (\ref{eq48}),
\begin{equation}
\zeta^+(s)=\frac{\pi^2}{(2\pi)^4}\frac{1}{\Gamma(s)}{\sum_{n=-\infty}^{\infty}}'\int_0^\infty \frac{dx}{\xi n^2} e^{-x}\left(\frac{x}{\xi n^2}\right)^{s-3}
\label{eq50}.
\end{equation}

We can express the $t$ integral in terms of the Gamma function
\begin{equation}
\zeta^+(s)=\frac{\xi^{2-s} \pi^2}{(2\pi)^4}\frac{\Gamma(s-2)}{\Gamma(s)}   {\sum_{n=-\infty}^{\infty}}'\frac{1}{n^{2s-4}}.
\label{eq51}
\end{equation}
We immediately recognise the infinite sum as the Riemann Zeta function so we finally obtain
\begin{equation}
\zeta^+(s)=\frac{\xi^{2-s} \pi^2}{(2\pi)^4}\frac{\Gamma(s-2)}{\Gamma(s)}   \zeta(2s-4).
\label{eq52}
\end{equation}

After expressing the gamma functions as 
\begin{equation}
\frac{\Gamma(s-2)}{\Gamma(s)}=\frac{\Gamma(s-2)}{(s-2)(s-1)\Gamma(s-2)}, 
\label{eq53}
\end{equation}
and plugging back into\ (\ref{eq48}), and performing the derivative with respect to s evaluated at $s=0$ we obtain
\begin{equation}
E=-\frac{\pi^2}{2\pi^4} \left(    \frac{(1+\alpha^2_\phi)^2\pi^2}{R^2}\right)^2\zeta'(-4).
\label{eq54}
\end{equation}
However, the derivative of the zeta function is known to be
\begin{equation}
\zeta'(-4)=\frac{3}{4\pi^4}\zeta(5),
\label{eq55}
\end{equation}
and so we find our final expression for the Casmir energy in an $S^1\times Z_2$ orbifold for a scalar field with periodic boundary conditions coupled to a Lorentz violating vector field to be
\begin{equation}
E=-\frac{3(1+\alpha^2_\phi)^2}{64\pi^2}\frac{1}{R^4}\zeta(5).
\label{eq56}
\end{equation}

We thus find that the Casimir energy contribution of a scalar field with periodic boundary conditons interacting with a Lorentz violating vector field remains attractive and tends to shrink the extra dimension. Thus, stabilization is not achieved with only scalars interacting with a Lorentz violating vector field. Note however that the expression for the effective potential takes into account the Casimir energy contribution from the bulk, but is incomplete because there can be additional contributions from the branes and other possible fields.

\section{Discussion}

In the context of radius stabilization, we have calculated the one loop corrections arising from a scalar field with periodic boundary conditions interacting with a Lorentz violating vector field in the compactified extra dimension of the Randall-Sundrum spacetime. The compactification scheme appears with enhanced sensitivity to the presence of periodic scalars interacting with Lorentz violating vectors (and tensors in general). In particular the contributions are attractive, inducing the extra dimension to shrink in size. Thus, a net positive contribution to the Casimir force from branes and additional fields is required for stabilization.

\section{Acknowledgements}

R.O. thanks Tibra Ali for helpful discussions. R.O would also like to thank Sean Carroll for clarifying aspects of the $Z_2$ orbifolding, and Masato Ito for explaining some of the subtleties of dimensional regularization. Research funding leading to this manuscript was partially provided by Baylor URC grant 0301533BP.

\end{document}